\documentclass{article}

\usepackage{rotating}
\usepackage{epsfig,amssymb,amsmath}

\def\to{\rightarrow}  \def\ev{\mbox{eV}}
  
\def\mpc{\mbox{Mpc}}

\def\al{\alpha} \def\be{\beta} \def\ga{\gamma} \def\de{\delta}
\def\ep{\epsilon}   
\def\th{\theta}   \def\ka{\kappa}
   
\def\si{\sigma}   
   
\def\om{\omega} \def\Ga{\Gamma} \def\De{\Delta} 
\def\La{\Lambda} \def\Si{\Sigma}  
  \def\mn{{\mu\nu}}

 \def\frac#1#2{{\textstyle{{#1}\over
{#2}}}} 
\def\lsim{\mathrel{\rlap{\lower4pt\hbox{\hskip1pt$\sim$}}
\raise1pt\hbox{$<$}}}
\def\gsim{\mathrel{\rlap{\lower4pt\hbox{\hskip1pt$\sim$}}
\raise1pt\hbox{$>$}}} \def\sqr#1#2{{\vcenter{\vbox{\hrule height.#2pt
\hbox{\vrule width.#2pt height#1pt \kern#1pt \vrule width.#2pt} \hrule
height.#2pt}}}}

\def\beq{\begin{equation}} \def\eeq{\end{equation}}
\def\beqa{\begin{eqnarray}} \def\eeqa{\end{eqnarray}}
\def\eq#1{Eq. (\ref{#1})}

\begin{document}

\title{The experimental status of Special and General Relativity}
\author{O. Bertolami and J. P\'aramos}

\maketitle 
\begin{abstract}
\noindent In this contribution we assess the current experimental status of Special and General Relativity. Particular emphasis is put on putative extensions of these theories and on how these could be detected experimentally.
\end{abstract}

\section{Introduction}

Special Relativity (SR) was proposed more than a hundred years ago and has allowed for a profound change in our perspective of the fundamental building blocks of physics, namely space and time, which were till then regarded as immutable and absolute. 

The generalization of SR to encompass general coordinate transformations and, through the Equivalence Principle, to also incorporate gravity, has lead to an inevitable connection to the mathematics of curved spaces, putting General Relativity (GR) in an unique standing among physical theories; GR is a theory of space and time, thus setting the tools to describe the dynamics and the evolution of the Universe as a whole.

From the conceptual point of view, Relativity was a major step forward; the pressure to unravel putative extensions to this theory of Gravity leads one to carefully test the foundational principles of SR and GR (see Refs. \cite{Will1,Will2005,survive} and references therein).

\section{Experimental Tests Special Relativity}
\label{exptestsSR}

More than a century ago, Einstein put forward his revolutionary special theory of Relativity (SR), so called because it accounted only for phenomenon seen from inertial reference frames, which move with constant relative velocity. Although several reformulations have arisen in the intervening years, with added clarity and mathematical precision \cite{Hodgson}, Einstein resorted to two fundamental postulates in order to derive SR:

\begin{itemize} 
\item The Principle of Relativity, which states that physical laws are independent of the inertial reference frame used to infer them,
\item The constancy of the speed of light, which is always propagated in empty space at $c \approx  3\times 10^8~m/s$, independently of the state of motion of its source. 
\end{itemize}

Both postulates may be shown to lead to the concept of Lorentz invariance, {\it i.e.} that the laws of physics are invariant with respect to the Lorentz transformations: if one takes two inertial frames $S$ and $S'$ with relative speed $v$ in the $x$-axis, these amount to the well known relations between time and space coordinates:

\beq x' = {x - vt \over \sqrt{1 - {v^2 \over c^2}}} ~~~~,~~~~y'=y~~~~,~~~~z'=z~~~~,~~~~ t' = {t - {v x \over c^2} \over \sqrt{ 1 - {v^2 \over c^2}}} ~~. \label{LT} \eeq

These transformations were known to leave Maxwell's equations invariant, while the Galileo transformations, which leave mechanics invariant, did not. However, it was Einstein who first understood that they could not be framed in a classical, Newtonian world-view, accompanied by a suitable aether medium, but instead required a fundamental rethinking of the concepts of space and time. The eponymous experiments carried out by Michelson and Morley in 1887 where taken not only as a disproof of this ``luminous aether'', but as an observational evidence for the constancy of the speed of light.

Notice that the above postulates do not make any claim concerning the equivalence between mass and energy, since they are of a kinematic (or geometric) nature alone. However, Einstein's derivation of this relationship resorts to a putative Lorentz invariance, used to establish the Lorentz invariant $(cp)^2 = E^2 - m^2c^4$ involving momentum, energy and rest mass.

The above introduction serve not only as an historical introduction, but helps to assess what are the most likely signals of Lorentz violation: privileged frame effects, variations in the speed of light \cite{VLT} or failure of the Lorentz transformations altogether; other consequences include deviations from the $p^2 = E^2 - m^2c^4$ dispersion law, or maximum attainable speeds $c_i \neq c$ for different matter species.

In the realm of theories of gravity, competing theories to the {\it de facto} gold standard of General Relativity (GR) are usually experimentally assessed via the so-called PPN formalism \cite{Will1}, discussed later in this text: for now, it suffices to state that this formalism relies on an expansion of the dynamical metric field $g_\mn$ in terms of suitable potentials, and the ensuing identification of a set of PPN parameters signaling deviation from GR. Since SR is restricted to inertial frames and negligible gravitational fields, thus assuming the {\it a priori} Minkowski metric $g_\mn = \eta_\mn$, such tool is not valid when addressing the issue of Lorentz symmetry breaking in SR.

Nevertheless, one may resort to a similar expansion of some fundamental relation or quantity, with the expansion coefficients being related to alternative theories to SR that breaks Lorentz invariance. The brief discussion above serves to better settle the three candidates that arise prominently: the Lorentz transformations themselves, the speed of light $c$ and the dispersion relation $p^2= p^2(E,m,\chi)$ (where $\chi$ symbolizes additional properties or fields not present in the equivalent SR relation).

Each of this ``test subjects'' leads to a widely different formalism, which also reflects whether its motivation is 

\begin{itemize}
\item[-] Kinematic, {\it i.e.} a relatively straightforward description of deviations from SR in the motion of massive bodies, propagation of light, causality, observability, light cone considerations, {\it etc.};
\item[-] Dynamical, in which case it attempts to formulate the intrinsic behaviour of fields and fundamental equations in terms of Lorentz breaking quantities, thus allowing one to capture other relativistic behaviour such as the clock rates of physical clocks ({\it e.g.} atomic clocks), light polarization effects, {\it etc}.
\end{itemize}

Given the variety of kinematic and dynamical  formalisms available, it is somewhat difficult to compare directly, either in terms of constraints to their defining observables, or when attempting to address a particular theoretical Lorentz breaking construction (see Ref. \cite{Lammerzahl} for a discussion).

\subsection{The Robertson-Sexl-Mansouri formalism}

Historically, this was the first attempt to put forward a formalism embodying the possibility of Lorentz symmetry breaking through the deviation of some free parameters from their SR values \cite{RSM}; this is addressed by assuming that a privileged frame $\Si (T,\vec{X})$ exists (usually considered the cosmological frame, defined as that where the cosmic microwave background radiation appears isotropic and homogeneous at large scales), so that transforming from this to another inertial frame $S(t,\vec{x})$ with relative velocity $\vec{v}$ is achieved via the deformed transformations:

\beq T = {t - {\ep} \cdot \vec{x} \over a}~~~~,~~~~ \vec{X} = {\vec{x} \over d} - \left({1 \over d} - {1 \over b}\right) {\vec{v}\cdot \vec{x}\over v^2} \vec{v} + \vec{v} T \eeq

Comparing with \eq{LT}, one finds that in SR,

\beq a = b^{-1} = \sqrt{1 - {v^2 \over c^2}} ~~~~,~~~~~d=1~~. \label{RSMSR} \eeq

The vector $\vec{\ep}$, although not uniquely determined, does not add any further information concerning a putative breaking of Lorentz invariance, but reflects the chosen clock synchronization procedure: momentary external synchronization leads to $\vec{\ep} = \vec{0}$, while Einstein synchronization implies $\vec{\ep} = -a \vec{v}/b(1-v^2)$; slow transport of clocks leads to $ ( \nabla_{\vec{v}} a) / b$, although additional complications due to clock rate variations between spacetime points, can arise from dynamical effects on the clock mechanism ({\it i.e.} shifts in atomic transition frequencies). As a result, physical observables do not depend on the particular choice of $\vec{\ep}$, with the natural exception of those experiments which directly depend upon a particular synchronization method (see Ref. \cite{Will1} for a discussion).

It is advantageous to resort to a set of numerical coefficient to parameterize the extend of violation of Lorentz symmetry, instead of using the functional form of $a$ and $b$: since most conceivable experiments involve massive bodies endowed with non-relativistic speeds, one may attain this by expanding these quantities around $(v/c)^2 $ to second-order, thus obtaining  

\beqa a(v) &\approx & 1 + \left(\al - {1 \over 2} \right) \left({v \over c}\right)^2 + \left(\al_2 - {1 \over 8} \right) \left({v \over c}\right)^4 ~~, \\ \nonumber b(v) &\approx & 1 + \left(\be + {1 \over 2}\right) \left({v \over c}\right)^2 + \left(\be_2 + {3 \over 8}\right)\left({v \over c}\right)^4 ~~, \\ \nonumber d(v) &\approx & 1 + \de \left({v \over c}\right)^2 + \de_2 \left({v \over c}\right)^4 ~~, \\ \nonumber \vec{\ep} &\approx & (\ep -1) \left[ 1 + \ep_2 \left({v \over c}\right)^2 \right] \vec{v} ~~.\eeqa

\noindent The choice of expansion coefficients is made so that, upon comparison with \eq{RSMSR}, one finds that SR yields all vanishing parameters except $\ep$ and $\ep_2$, as discussed above; Einstein synchronization, the usual procedure followed in SR, also yields $\ep = \ep_2 = 0$. 

Using the above expressions, one may derive a convoluted expression for the speed of light,

\beqa c &=& 1 - \ep \cos\th {v \over c} - \bigg[ \de - \al + (\be - \ga + \ep^2) \cos^2 \th \bigg] \left({v\over c}\right)^2 +  \\ \nonumber && \bigg[ \be - \al + \ep_2 - \ep (2 [\al + \de] + \ep_2 ) - \ep (2[\be - \de] + \ep^2 ) \cos^2 \th \bigg] \cos \th \left({v\over c}\right)^3 + \\ \nonumber && \bigg[ \de_2 - \al_2 -\al\left( {1\over 2} + \de - \al \right) + \bigg(\be_2 - \de_2 - \be \left[{1 + 3 \be \over 2} + \al -3\de + 2\ep \right] + \\ \nonumber && + \al [\de + (2-3\ep)\ep] - 3\de \left[ {\de \over 2} - \ep^2 \right] + 2 [\ep-1]\ep\ep_2 + \\ \nonumber && \left[ 3 (\be - \de) \left( {\be - \de \over 2} + \ep^2 \right) + \ep^4  \right] \cos^2 \th \bigg) \cos^2 \th \bigg] \left({v\over c}\right)^4~~. \eeqa

\noindent where $\th$ is the angle between the velocity ${\bf v}$ of the frame of reference and the path of light; the independence of experiments not relying on a specific synchronization method on the related parameters $\ep$ and $\ep_2$ becomes apparent if one computes the relative shift in the two-way speed of light $c_2(\th,v)$,

\beqa {c_2(\th,v) \over c_2(0,v)} -1 &=& \sin^2\th \bigg[ (\de - \be) \left({v\over c}\right)^2 + \\ \nonumber && \left( {3 \de^2 - \be^2 \over 4} + \be_2 - {\be \over 2}(1+\de) - \de_2 + {3 \over 4}(\be-\de)^2 \cos 2\th \right) \left({v\over c}\right)^4 \bigg]~~.\eeqa

\noindent Similarly, the phase shift (not shown here for brevity), which can be measured by interferometry, is also independent on $\ep$ and $\ep_2$ (see Ref. \cite{Lammerzahl} for details).

An experimental determination of any non-vanishing parameters would immediately indicate that the underlying physical theory is not Lorentz invariant. However, second-order tests ({\it i.e.} obtained by disregarding terms $O(v^4)$ above) have yielded impressive bounds on these quantities: the most recent Michelson-Morley experiment probing the dependence of the speed of light on its orientation with respect to a preferred frame has yielded $(\be-\de) = (4 \pm 8)\times 10^{-12}$ \cite{Herrmann}, while a modification of its setup (the Kennedy-Thorndike experiment) has shown no signal of an effect of $c$ on the velocity of the apparatus, $(\al - \be ) = -4.8 (3.7) \times 10^{-8} $ \cite{Tobar}; finally, the most precise relativistic Doppler effect measurement has shown that time dilation as predicted by SR is valid down to a precision of $|\al | \leq 8.4 \times 10^{-8} $ \cite{Reinhardt}. Thus, no violation of SR or any of its foundational principles has been detected so far.

\subsection{The $c^2$ formalism}

Another formalism to address the possibility of breaking Lorentz invariance arises if one disregards the postulate of the constancy of the speed of light\footnote{For clarity, one does not assume in this paragraph the natural system of units, in which $c = 1$.}. This is best attained by resorting to the so-called $TH\ep \mu$ framework \cite{Lightman}, an alternative to the PPN formalism when parameterizing gravity theories that deviate from GR \cite{Will1}.

This formalism is well-suited to describe the interaction between charged particles in an external static and spherically symmetric gravitational field resulting from some metric theory of gravity: the field $T = g_{00} $ describes the temporal component of the metric $g_\mn$, while isotropy allows one to express the spatial part as $g_{ij} = H \eta_{ij}$. The $\mu$ and $\ep$ parameters act as a generalization of the magnetic permeability $\mu_0$ and electric permitivity $\ep_0$ of a medium: depending on the underlying physical theory, these may depend on internal structure or the effect of other bodies.

The $TH\ep\mu$ formalism is able to signal deviations from metricity via a set of appropriately defined parameters

\beqa \Ga_0 = -c_0^2 {\partial \over \partial U} \ln \left[ \ep \sqrt{T \over H}\right] ~~, \\ \nonumber \La_0 = -c_0^2 {\partial \over \partial U} \ln \left[ \mu \sqrt{T \over H}\right]~~, \\ \nonumber \Upsilon_0 = 1 - {T \over H} \ep \mu~~  , \eeqa

\noindent which vanish if the EP is valid\footnote{More rigorously, the Einstein Equivalence Principle (EP), which comprises the Weak EP, Local Lorentz invariance and Local Position Invariance.}. In the above, $c_0 = \sqrt{T /H}$ is shown to be the limiting speed of material test particles; the latter contrasts with the speed of light $c = 1/\sqrt{\ep\mu}$, which follows the usual definition stemming from Maxwell's equations; however, both speeds $c_0$ and $c$ allow for a spacetime and/or constitution dependency, that is, a non-constant $c$ breaks Einstein's second postulate of SR, while the Equivalence Principle  is broken if $c \neq c_0$ (even if $c_0 = {\mathrm const.}$ is the same for all matter species).

Since SR is obtained by taking the flat spacetime limit of GR, one may extract a suitable formalism to address Lorentz symmetry breaking in negligible gravitational fields and inertial frames by considering the same limiting case of the $TH\ep\mu$ formalism: this is achieved by considering the $c^2$ formalism \cite{Will1}, and is attained by removing the spacetime dependence of the eponymous set of parameters, as if the dynamics of the gravitational field are disregarded. As a result, one is left with the possibility of deviations between $c$ and $c_0$.
 
\subsection{Modified dispersion relation} 

A more phenomenological, straightforward way of breaking Lorentz invariance is to assume that the SR dispersion relation $E^2 = p^2c^2 + m^2 c^4$ is generalized to $E^2 = F(p,E)$, due to some underlying physical theory. Knowing the latter, one should also be able to establish the conservation laws for energy and momentum: in the absence of full knowledge of its inner workings, one may assume that both quantities are conserved, or resort to another phenomenological dependency for $\De E(p,E)$ and $\De p(p,E)$.

Since SR has withstood all  tests so far, one knows that its dispersion relation must be a very good approximation, at least for the experimental regime available $v \ll c$. Thus, it is natural that the putative full dependence $E^2 = F(p,E)$ allows a Taylor expansion around $v = 0$, of the form

\beq E^2 = m^2 + p^2 + M_P f_i^{(1)}p^i + f_{ij}^{(2)}p^ip^j  + {f_{ijk}^{(3)} \over M_P}p^i p^j p^k + ...  ~~, \eeq

\noindent setting $c=1$, for simplicity; the coefficients $f^{(n)}$ are dimensionless, having factored out the Planck mass $M_P$, the assumed scale at which relevant Lorentz symmetry breaking effects should arise due to some fundamental theory. These coefficients must be related to the underlying physical theory, and could be space-time or position dependent. More evolved modifications of the dispersion relation may arise if one assumes that spacetime is discretized \cite{Dowker,JackNg} or stochastic \cite{Shiokawa}.

\subsection{Dynamical framework}

The previous formalisms address the phenomenological implications of breaking Lorentz invariance via deformed relations for the coordinate transformations, the dispersion relation or the speed of light or limiting velocity of massive bodies. In stark contrast, one may conceive dynamical schemes that attempt to model Lorentz breaking extensions via an effective theory, valid at the low-energy, low-velocity regime.

Since one is dealing with the issue of testing Lorentz invariance at low-energies, {\it i.e.} probing the validity of SR, gravity may be discarded from such an extension. The first set up in flat space is the minimal Standard Model extension (mSME) \cite{mSME}; in this, the interactions of the Standard model are enriched by a set of renormalizable Lorentz breaking operators involving fermions and the gauge bosons compatible with the internal gauge symmetry of QED. One may readily extend this to Yang-Mills theories, including models of the electromagnetic, weak and strong forces with an appropriate covering group. Gravity can also be included, as well as an embedding of our worldsheet into higher-dimensional braneworlds. Naturally, this dynamical framework encompasses the previously considered Robertson-Sexl-Mansouri formalism \cite{Mewes2}.

One focuses the attention on the ``minimal QED extension'', as it provides a sufficiently broad framework for the bulk of experimental tests of SR that involve electrons and light propagation. Further imposing the $SU(2)$ gauge symmetry breaking, one can write the relevant additional terms to the Lagrangean density describing fermions and the electromagnetic field as \cite{Mewes}

\beq \Delta {\cal L} = {1 \over 2} \bar{\psi} \Ga_\nu \stackrel{\leftrightarrow}{\partial^\nu}\psi -\bar{\psi}M\psi + {1 \over 2} (\ka_F)_{\al\be\mn}F^{\al\be}F^\mn~~, \eeq

\noindent where $F_\mn = \partial_\mu A_\mu - \partial_\nu A_\mu$ is the usual field strength tensor. In the fermionic sector, one introduces a generalized mass term

\beq M \equiv m + a_\mu \ga^\mu + b_\mu \ga_5 \ga^\mu + {1 \over 2} H_\mn \si^\mn~~, \eeq

\noindent where $m$ is the ``bare'' mass, as well as generalized gamma matrices 

\beq \Ga_\nu \equiv \ga_\nu + c_\mn \ga^\mu + d_\mn \ga_5 \ga^\mu + e_\nu + i f_\nu \ga_5 + {1 \over 2} g_{\al \mn} \si^{\al \mu}~~, \eeq 

\noindent where the $a_\mu$, $b_\mu$, $c_\mn$, $d_\mn$, $e_\nu$, $f_\nu$, $\si_\mn$ and $H_\mn$ are parameters that should arise from the underlying high energy theory. It is worthwhile to notice that if the breaking of Lorentz invariance is spontaneous, {\it i.e.} this is an exact symmetry of the latter, then these parameters are related to vacuum expectation values of Lorentz tensors and must be CPT invariant \footnote{A toy model where a suitable number of vectors couple to the Ricci curvature introduces, when the former acquire a vacuum expectation value, a spontaneous Lorentz symmetry breaking into the gravity sector, and yields interesting astrophysical implications \cite{bumblebee,flight}.}. Hermiticity of ${\cal L}$ also implies that they are real.

Dropping higher-order operators, which should not be as relevant in the low-energy limit, one expects a fermionic (odd) term of the form $(\ka_{AF})^\al \ep_{\al\be \mn} A^\be F^\mn $; however, since this gives rise to negative contributions to the canonical energy and may lead to instabilities in the theory \cite{Jackiw,PerezVictoria}, it is usually considered to vanish, $k_{AF} = 0$ ---  which is experimentally supported. 

Given the suggestive notation above, one naturally obtains a generalized Dirac equation

\beq (i \Ga^\mu \partial_\mu - M)\psi = 0~~, \eeq

\noindent together with generalized inhomogeneous Maxwell equations (without sources), 

\beq \partial_\nu F^{\mu\nu} + {(\ka_F)^\mu}_{\nu\al\be} \partial^\nu F^{\al\be} = 0~~, \eeq

\noindent while the homogeneous Maxwell equations remain the same. The full set may be suggestively recast into the usual counterpart, $\partial_\mu F^\mn = 0$, but with the deformed constitutive relations for the medium,

\beq  \left[\begin{array}{c}\vec{D} \\\vec{H}\end{array}\right] = \left [\begin{array}{cc}\ep_0 (\stackrel{\sim}{\ep}_r + \ka_{DE}) & \sqrt{\ep_0 \over \mu_0} \ka_{DB} \\ \sqrt{\ep_0 \over \mu_0} \ka_{HE} & \mu_0^{-1} (\stackrel{\sim}{\mu}_r^{-1} + \ka_{HB}) \end{array}\right] \left[\begin{array}{c}\vec{E} \\\vec{B}\end{array}\right] ~~, \eeq

\noindent where $\stackrel{\sim}{\ep}_r$, $\stackrel{\sim}{\mu}_r$ are the electric permitivity and magnetic permeability matrices, respectively, (proportional to the $3 \times 3$ identity matrix for linear, homogeneous and isotropic mediums) and one defines

\beqa {\ka_{DE}}^{ij} &=& -2 {\ka_F}^{0i0j}~~~~,~~~~  {\ka_{HB}}^{ij} = {1 \over 2} \ep^{ikl} \ep^{jmn}{\ka_F}^{klmn}~~, \\ \nonumber  {\ka_{DB}}^{ij} &=& - {\ka_{HE}}^{ji} = {\ka_F}^{0ikl} \ep^{jkl}~~. \eeqa

Hence, one has the ingredients to perform a thorough analysis of a possible breaking of Lorentz invariance involving charged particles and light.

In four spacetime dimensions, renormalizability of Standard model operators requires that these have a mass dimension $d \leq 4$; however, in principle, Lorentz breaking operators with any mass dimension $d$ could also appear in the Lagrangean of the effective field theory extending the Standard model at low-energies.

If the lower-dimensional operators $d \leq 4$ are not adequately suppressed at low-energy scales, they dominate the higher-dimensional ones and lead to unacceptably high corrections to the deformed dispersion relation and have the form $f^{(n)} p^n M_P^{2-n}$, with $f^{(n)} \sim 1$. Moreover, radiative corrections lead to additional linear and quadratic terms of the form $M_P p + f^{(n}) p^2$.

Since it is known experimentally that the dispersion relation of SR holds with great accuracy (see Section \ref{LLI}), one requires either an unnatural fine-tuning of the dimensionless coefficients affecting these operators \cite{Collins}, so that additional linear and quadratic terms in the deformed dispersion relation cancel out. An explicit computation of the dispersion relation from the mSME for fermions is found in Ref. \cite{Bertolami2}.

As it turns out, one may resort to partial discrete symmetries that remain after the main one is broken: a natural candidate is $CPT$, as the odd Lorentz invariant operators of the $mSME$ are restricted (and thus made compatible with the experimental bounds) if one enforces this symmetry in the theory. The even operators may also be suppressed if one invokes supersymmetry as a natural invariance of Nature, although consistency requires that allowed Lorentz symmetry breaking operators involving supersymmetric partners are also considered, and even operators remain dangerously unrestricted.

The kinetic and dynamic formalisms presented above are all naturally intertwined, and may be correlated although the underlying physical theory remains unknown --- that is, a particular Lorentz breaking contribution to the effective field theory envisaged in the dynamical framework naturally translates into specific modified dispersion relations \cite{Myers,Bertolami:2004bf}, while phenomenological terms considered in the kinetic approach can in principle be traced back to relevant operators at the low-energy level \cite{Mewes}. 

\section{Testing General Relativity}
\label{sec:gr-survey}

Having discussed above how the foundational principles of Special Relativity can be tested, one now focuses on GR and its current experimental status. The first experimental confirmation of GR appeared in 1915, when it successfully accounted for the discrepancy with the Newtonian estimate for the advance of the perihelion precession of Mercury's orbit with no adjustable parameters. Shortly after, the famous 1919 expedition by Eddington produced observations of stellar lines-of-sight during a solar eclipse that confirmed another prediction of GR, namely that the deflection angles due to light bending around the gravitational field of the Sun should be twice the value obtained from Newtonian and Equivalence Principle arguments. This propelled GR into notoriety and turned its creator into the first scientific star of the world.

Since then, GR has been extensively tested in the Solar System, with all data obtained so far being consistent with its predictions. As time went by, these tests have grown more and more precise: from the $\sim 0.2$ accuracy of microwave ranging to the Viking Lander on Mars in 1976 \cite{viking_shapiro1,viking_reasen,viking_shapiro2} and $0.15$ accuracy of spacecraft and planetary radar observations \cite{anderson02} to an order of magnitude gain via the astrometric observations of quasars on the solar background performed with Very-Long Baseline Interferometry \cite{RoberstonCarter91,Lebach95,Shapiro_SS_etal_2004} and lunar laser ranging precision measurements of the lunar orbit (with accuracies of $\sim 0.045$ and $\sim 0.011 $, respectively) \cite{Ken_LLR68,Ken_LLR91,Ken_LLR30years99,Ken_LLR_PPNprobe03,JimSkipJean96,Williams_etal_2001,Williams_Turyshev_Boggs_2004}. This was pushed even further by the 2003 experiments with the Cassini spacecraft, which improved the testing accuracy down to $\sim 0.0023 $ \cite{cassini_ber}.

Observations of binary millisecond pulsars lend further support for GR: indeed, the physical processes occurring in the strong gravitational field regime within these relativistic object are of considerable interest, given the possibility of testing relativity in a distinct dynamical environment. Pulsar tests of strong-field gravity were first formulated in Refs. \cite{DamourTaylor92}, with initial tests being performed with PSR1534 \cite{Taylor_etal92}. Strong-field gravitational tests and their theoretical rationale was examined in Ref. \cite{Damour_EFarese96a,Damour_EFarese96b,Damour_EFarese98}. Pulsar data were recently analyzed to test GR to $\sim 0.04 $ at a $3 \sigma$ confidence level \cite{lange_etal2001}.

\subsection{Metric Theories of Gravity and PPN Formalism}

In this section, one presents the formalism used to interpret observations in the weak-field and slow motion approximation, conditions found in the Solar System; this formalism provides a rigorous framework to study increasingly accurate experiments and to establish stringent constraints on deviations from GR and its fundamental tenets.

The distribution of matter in this approximation is commonly represented by a perfect fluid model \cite{Fock1,Fock1a,Fock2,Chandrasekhar_65} with an energy-momentum tensor $\widehat{T}^{mn}$ given by

\beq \widehat{T}^{mn}=\sqrt{-g}\left(\left[\rho_{0}( 1 + \Pi) + p\right]u^{m} u^{n} - p g^{mn} \right)~~, \label{eq:perfect-fl} \eeq

\noindent where $\rho_0$ is the mass density of the ideal fluid in coordinates of the co-moving frame of reference, $u^k $ are the components of invariant four-velocity of a fluid element, and $p(\rho)$ is the isentropic pressure connected with the energy density by an equation of state $p = p(\rho)$. The quantity $\rho\Pi$ is the density of internal energy of an ideal fluid; the definition of $\Pi$ arises from the first law of thermodynamics, according to the equation $ u^n\big(\Pi_{;n} + p\big({1/ {\widehat \rho}}\big)_{;n}\big) = 0 $, where ${\widehat \rho}=\sqrt{-g}\rho_0u^0$ is the conserved mass density \cite{Fock2,Chandrasekhar_65,Brumberg,Will1}. Considering the energy-momentum tensor, the solutions of the gravitational field equations for a given theory of gravity can be found.

An alternative methodology, valid for both the weak and strong regimes of GR and an arbitrary energy-stress tensor, builds upon a ``Maxwell-like'' expansion of the metric and the Blanchet-Damour multipole framework \cite{DSX1,DSX2,DSX3,Blanchet_etal_95,Damour_Vokrouhlicky_95}; the study of a general N-body problem in a weak-field and slow motion approximation was developed in Ref.~\cite{Kopeikin_Vlasov_2004}.

Despite the widely different principles underlying metric theories of gravity, they all share the feature that the gravitational field directly affects the matter through the metric tensor $g_{mn}$, which is determined from the field equations. Thus, the metric expresses the properties of a particular gravitational theory and carries information about the bodies' gravitational field --- contrasting with the flat metric of Newtonian gravity and its interpretation in terms of forces acting at a distance.

The so-called parameterized post-Newtonian (PPN) formalism generalizes the phenomenological parameterization of the gravitational metric tensor field first discussed by Eddington in a limited context \cite{Nordtvedt_1968a,Will_1971,Will_Nordtvedt_1972}. This method assumes slowly moving bodies and weak inter-body gravity, and is valid for a broad class of metric theories. The PPN parameters that appear in the expansion of the metric characterize each theory of gravity, and are individually associated with the underlying symmetries and laws of invariance. If, for simplicity, one assumes Lorentz and Local Position Invariance and conservation of total momentum conservation, the metric tensor in four dimensions in the PPN-gauge is given by
\beqa 
        g_{00} &=&
        -1 + 2U - 2 \beta U^2 - 2 \xi \Phi_W
        + (2 \gamma +2+ \alpha_3 + \zeta_1 - 2 \xi ) \Phi_1 \nonumber \\ \nonumber
        &&+ 2(3 \gamma - 2 \beta + 1 + \zeta_2 + \xi ) \Phi_2
        + 2(1 + \zeta_3 ) \Phi_3
        + 2(3 \gamma + 3 \zeta_4 - 2 \xi ) \Phi_4 \\ \nonumber
        &&- ( \zeta_1 - 2 \xi ) {\cal A}
        - ( \alpha_1 - \alpha_2 - \alpha_3 ) w^2 U
        - \alpha_2 w^i w^j U_{ij}
        + (2 \alpha_3 - \alpha_1 ) w^i V_i \\ \nonumber
        && +  O(\epsilon^3)~~, \\ \nonumber
        g_{0i}
        &=& - {1 \over 2 }
        (4 \gamma + 3 + \alpha_1 - \alpha_2
        + \zeta_1 - 2 \xi ) V_i
        - {1 \over 2}
        (1 + \alpha_2 - \zeta_1 + 2 \xi )W_i \\ \nonumber
        &&- {1 \over 2} ( \alpha_1 - 2 \alpha_2 ) w^i U
        - \alpha_2 w^j U_{ij} + O(\epsilon^{5/2})~~, \\
        g_{ij}
        &=& (1 + 2 \gamma U ) \delta_{ij} + O(\epsilon^2)~~,\label{eqno(1)}
      \eeqa

\noindent setting $\hbar=c=G=1$ and using the metric signature convention $(-+++)$.

The order of magnitude of the various terms is determined according to the estimates $U \sim v^2 \sim \Pi \sim p/\rho
\sim \epsilon$, $v^i \sim |d/dt|/|d/dx| \sim \epsilon^{1/2}$, and all possible potentials are considered up to the desired Post-Newtonian order. Considering Eq.~(\ref{eq:perfect-fl}), these generalized gravitational potentials, of the same order as $ U^2 $, are given by

  \beqa U&=&\int {{\rho' } \over {| {\bf x}-{\bf x}' |}}        d^3x'~~, \nonumber \\ \nonumber        U_{ij}&=&        \int {{\rho' (x-x')_i (x-x')_j}          \over {| {\bf x}-{\bf x}' |^3}} d^3x'~~, \\\nonumber        \Phi_W &=&        \int {{\rho' \rho'' ({\bf x}-{\bf x}')}  \over {| {\bf x}-{\bf x}' |^3}} \cdot \left(          {{{\bf x}' -{\bf x}'' }            \over {| {\bf x}-{\bf x}'' |}}-          {{{\bf x}-{\bf x}''}
            \over {| {\bf x}'-{\bf x}'' |}}        \right) d^3x' d^3x''~~, \\ \nonumber        {\cal A} &=& \int {{\rho' [{\bf v}'            \cdot ({\bf x}-{\bf x}')]^2 }          \over {| {\bf x}-{\bf x}' |^3}}        d^3x'~~, \\ \nonumber        \Phi_1 &=&        \int {{\rho' v'^2}          \over {| {\bf x}-{\bf x}' |}} d^3x'~~, \\ \nonumber        \Phi_2 &=&        \int {{\rho' U'} \over {| {\bf x}-{\bf x}' |}}        d^3x'~~, \\\nonumber        \Phi_3 &=&        \int {{\rho' \Pi'} \over {| {\bf x}-{\bf x}' |}}        d^3x'~~, \\ \nonumber        \Phi_4&=&        \int {{p' } \over {| {\bf x}-{\bf x}' |}} d^3x'~~, \\ \nonumber        V_i &=&        \int {{\rho' v_i'} \over {| {\bf x}-{\bf x}' |}} d^3x'~~, \\W_i&=&        \int {{\rho' [{\bf v}' \cdot            ({\bf x}-{\bf x}')](x-x')_i}          \over {| {\bf x}-{\bf x}' |^3}} d^3x'~~.\label{eq:gen-pots}      \eeqa%

A particular metric theory of gravity in the PPN formalism is fully characterized by means of the eleven PPN parameters shown in \eq{eqno(1)} \cite{Will1,Turyshev96}: these have clear physical meaning, and concern a particular symmetry, conservation law or fundamental tenet of the structure of spacetime: the parameter $\beta$ is the measure of the non-linearity of the law of superposition of the gravitational fields (or its metricity) in a theory of gravity, while $\gamma$ represents the measure of the curvature of the spacetime created per unit rest mass; the group of parameters $\alpha_1, \alpha_2, \alpha_3$ quantify the violation of Lorentz invariance ({\it i.e.} the existence of the privileged reference frame), the parameter $\zeta$ quantifies the violation of Local Position Invariance, and the parameters $\al_3,\zeta_1,\zeta_2,\zeta_3,\zeta_4$ indicate a possible violation of the conservation of total momentum. 

Since GR satisfies all of the above principles, it is naturally signaled by the vanishing of all PPN parameters except $\be$ and $\ga$. Brans-Dicke theory \cite{Brans}, perhaps the best known of the alternative theories of gravity, endowed with an additional scalar field and arbitrary coupling constant $\omega$, yields a decreasing spacetime curvature per unit rest mass, while preserving the remaining symmetries: its non-vanishing PPN parameters are thus $\beta=1$, $\gamma= ( 1 + \omega ) / ( 2 + \omega )$. More general scalar tensor theories yield values of $\beta$ different from unity \cite{Damour_Nordtvedt_1993a}.

\begin{figure}[t]
\centering
\leavevmode\epsfxsize=11.5cm \epsfbox{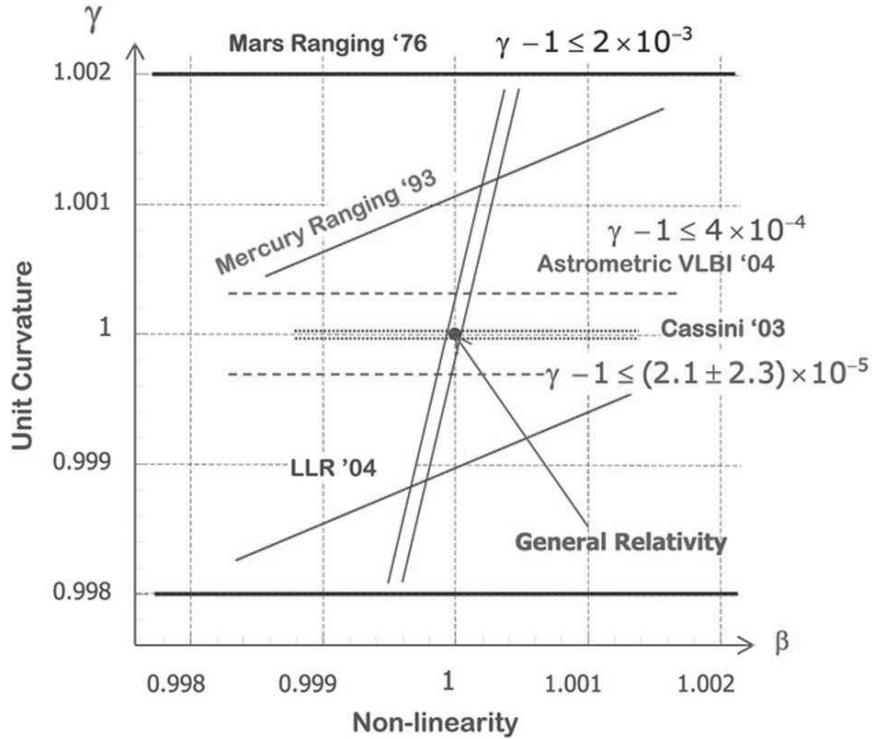}\\
\caption{\label{fig:Will1} The progress in determining the PPN
parameters $\gamma$ and $\beta$ for the last 30 years (adopted
from \cite{turyshev_acfc_2003}).}
\end{figure}

The PPN metric tensor, given by Eqs.~(\ref{eqno(1)}-\ref{eq:gen-pots}), is used to generate the equations of motion for the bodies under scrutiny (planets, satellites, {\it etc}.), which are translated into orbit determination numerical codes \cite{Moyer81a,Moyer81b,Standish_etal_92,Turyshev96}, as well as being used in the analysis of gravitational experiments in the Solar System \cite{Will1,turyshev_acfc_2003}. Table \ref{ppntable} and Fig. \ref{fig:Will1} show the latest bounds on the Eddington parameters $\be$ and $\ga$ and the history of increasingly accurate experiments.

\begin{table}[t!]
\caption{Accuracy of determination of the PPN parameters 
$\gamma$ and $\beta$ \cite{Williams_Turyshev_Boggs_2004,turyshev_acfc_2003,Will2005}.
\vspace*{0.5cm}}

\label{ppntable}

\begin{center}
\begin{tabular}{|c|c|c|}
\hline
  PPN parameter & Experiment & Result \\ \hline\hline
  $\ga -1 $ & Cassini 2003 spacecraft radio-tracking &
$ 2.3 \times 10^{-5}$ \\\cline{2-3}
  ~  & Observations of quasars with Astrometric VLBI & $ 3 \times 10^{-4} $ \\\hline 
  $ \be - 1 $ & Helioseismology bound on perihelion shift & $ 3 \times 10^{-3} $  \\\cline{2-3}
  ~ & LLR test of the SEP, assumed: $\eta = 4 \be - \ga - 3$   & $ 1.1 \times 10^{-4} $  \\
  ~ &  and the Cassini result for PPN $\gamma$    &  ~ \\
\hline

\end{tabular}
\\[10pt]
\end{center}

\end{table}

The foundations of GR and the current experimental verification of their validity are now discussed. For this, one recalls its basic tenets:

\begin{enumerate}

\item[1).]  Weak Equivalence Principle (WEP) (also known as the principle of universality of the free fall): freely falling bodies have the same acceleration in the same gravitational field, independently of their compositions (see Section~\ref{sec:eep});

\item[2).] Local Lorentz invariance (LLI): the rate of clocks is independent on the velocity of the clock (see Section~\ref{LLI});

\item[3).] Local position invariance (LPI): the rate of clocks is independent on the spacetime position of the clock (see Section~\ref{LPI}).

\end{enumerate}

\subsection{The Equivalence Principle (EP)}
\label{sec:eep}

Almost every theory of gravitation has addressed the issue concerning the equivalence between inertial and passive gravitational mass, starting with Newton himself. Almost one century ago, Einstein followed through by declaring that all non-gravitational laws should behave in free-falling frames as if gravity was absent. This postulate implies that identical accelerations should be experienced by objects with different compositions in the same gravitational field --- so that gravity becomes a geometrical property of spacetime, as posited by GR. As it turns out, this EP can be cast in both a weak and strong version, as addressed below.

\subsubsection{The Weak  Equivalence Principle (WEP)}
\label{sec:wep}

The weak form of the EP states that the gravitational properties of all interactions except gravity obey the EP. The concerned ``charges'' are the nuclear-binding energy differences between test masses, their neutron-to-proton ratios or atomic charges, amongst others. The equivalence between gravitational and inertial masses implies that distinct neutral massive test bodies have the same free fall acceleration in an external gravitational field \cite{Anderson_etal_1996}, with the latter inducing only a tidal force \cite{Singe_1960}.

According to GR, the spacetime curvature caused by a massive body scatters light rays passing in its vicinity achromatically. The Sun is the dominating contributor to this effect in the Solar System, deflecting the light by as much as $1.75'' \cdot (R_{\odot} / b)$, where $R_{\odot}$ is the solar radius and $b$ is the impact parameter. In 1919, the famous Eddington expedition confirmed that photons free fall according to the predictions of GR: although the original experiment had only a10\% accuracy, the light bending measured in a solar conjunction by the Cassini spacecraft has improved this type of measurement to the current figure of 0.0023\% \cite{cassini_ber}. 

The WEP also implies a Doppler frequency shift $\De \nu$ induced on light by the variation of the gravitational potential. This was confirmed in 1960 by the eponymous Pound-Rebka experiment, which produced
\beq {\Delta \nu \over \nu} = {g H \over c^2} = (2.57 \pm 0.26) \times 10^{-15}, \label{eq:1.15} \eeq \noindent
\noindent where $g$ is the acceleration of gravity and $H$ the height of fall \cite{Pound-Rebkaa,Pound-Rebkab}.

Notwithstanding some formidable experimental obstacles, the free fall of antiprotons and antihydrogen (or other antiparticles) could provide yet another test of the WEP (see Ref. \cite{Nieto} for a thorough review). This would allow one to probe to what extent does gravity respect the CPT symmetry of local quantum field theories --- specifically, if antiparticles fall as particles in a gravitational field. The ATHENA (ApparaTus for High precision Experiments on Neutral Antimatter) and the ATRAP collaborations at CERN have developed the capability of storing antiprotons and creating an antihydrogen atom \cite{Athena,Atrap}, but no experiment along these lines has been performed so far. 

A test of the WEP involving neutral kaons was performed by the CPLEAR collaboration \cite{Apostolakis}, producing limits of $6.5$, $4.3$ and $1.8 \times 10^{-9}$ respectively for scalar, vector and tensor potentials originating from the Sun with a range much greater than 1~AU acting on kaons and antikaons. These relevant results do not probe possible baryon number dependent interactions, and are thus complementary to the desirable antiprotons and antihydrogen atom experiments mentioned above.

Most metric theories of gravitation inherently uphold the WEP, although some predict additional forces that lead to composition-dependent deviations from geodesic motion ({\it e.g.} if a non-minimally coupling between matter and curvature is present \cite{Bertolami2007,BertolamiReview}). Similarly, almost all extensions to the standard model of particle physics predict new forces that induce apparent violations of the EP \cite{Damour_1996,Damour_2001}; this is most apparent if macroscopic-range fields are present, so that exchange forces that couple to generalized charges arise, rather than just to mass/energy as does gravity \cite{Damour_Polyakov_1994aa,Damour_Polyakov_1994ab}.

Laboratory tests of the WEP can be made by comparing the free fall accelerations $a_1$ and $a_2$, of different test bodies. If these are at the same distance from the source of the external gravitational field, the breaking of the WEP is elegantly gauged through the quantity
\beq {\Delta a \over a} = {2(a_1- a_2) \over a_1 + a_2} = \left({M_G \over M_I}\right)_{\hskip -3pt 1} -\left({M_G \over
M_I}\right)_{\hskip -3pt 2} = \Delta\left({M_G \over M_I}\right), \label{WEP_da} \eeq
\noindent where $M_G$ and $M_I$ are respectively the gravitational and inertial masses of each body.

Other tests conducted so far have validated the WEP for elementary particles. For the neutron, an interferometry experiment showed that a neutron beam split by a silicon crystal and traveling through different gravitational paths interferes as predicted by quantum mechanics, with a gravitational potential given by Newtonian gravity --- providing a striking confirmation of the WEP using an elementary hadron  \cite{Collela}. Since then, gravitational atom interferometric measurements have probed the WEP down to a precision of $3 \times 10^{-8}$ \cite{Kasevich}.

The ratio of gravitational to inertial masses of test bodies has been determined, with an upper limit for $| 1 - M_G/M_I |$ of $\sim 10^{-11}$ in 1964 \cite{Roll_etal_1964}, $\sim 10^{-12}$ in 1972 (reconfirmed in 1994) \cite{Braginsky_Panov_1972,Su_etal_1994} and, more recently, $1.4\times 10^{-13}$ \cite{Adelberger_2001} (see Ref. \cite{Gundlach} for a review). These increasingly precise experiments further show that the strong, weak, and electromagnetic interactions contribute equally to the passive gravitational and inertial masses of test bodies.

One decade ago, gravitational bound states of neutrons were confirmed by Nesvizhevsky and collaborators \cite{Nesvizhevsky,Nesvizhevsky1}, who set up a realization of a conceptual experiment proposed in 1978 \cite{Luschikov}. In this experiment, ultracold neutrons from a source at the Institute Laue-Langevin reactor in Grenoble fall under the influence of the Earth's gravitational field towards a horizontal mirror, with a minimum measurable energy of $1.4 \times 10^{-12}~\ev$ corresponding to a vertical velocity of $1.7$~cm/s (a more intense beam and an enclosure mirrored on all sides could lower the latter by six orders of magnitude). The neutrons were found not to fall continuously; rather, they jumped between different vertical levels, as predicted by quantum mechanics. 

Improved experiments probing gravity through quantum systems clearly open the possibility of testing novel concepts related to the unification of GR and quantum mechanics (in the low-energy regime), such as non-commutative formulations of the latter \cite{Bertolami18} --- as well as detecting the transition between the classical and quantum description of a system as a function of its dimensions \cite{Bertolami19}. 

An analysis of the lunar laser ranging data showed the absence of any composition-dependent acceleration effects \cite{Baessler_etal_1999}.
In astronomical measurements, one should consider the gravitational self-energy contributions to the inertial and gravitational masses of the bodies \cite{Nordtvedt_1968a}, whereas these are negligible in test masses used in laboratory environments. Considering the gravitational self-energy leads one to scrutinize the strong equivalence principle, as is discussed below.

\subsubsection{The Strong Equivalence Principle (SEP)}
\label{sec:sep}

The strong formulation of the EP addresses the gravitational behaviour of arising from gravitational energy itself, thus expressing the non-linearity of gravitation. It states that not only the outcome of gravitational experiments, but indeed any measurement concerning other interactions, are independent of the velocity and position of the laboratory. Being an integral part of the EP, the SEP is enforced by GR. However, many theories of gravity do not respect this assumption: for instance, scalar-tensor theories typically violate the SEP \cite{Nordtvedt_1968a,Nordtvedt_1968b,Ken_LLR68,Nordtvedt_1991}, {\it e.g.} by positing different couplings between these fields and different species of matter. This leads not only to a difference in free fall and related tests, but also on non-gravitational experiments.

The fractional contributions to the mass by gravitational self-energy of a body is the most relevant quantity for probing the validity of the SEP. The previously described PPN formalism is particularly suited to the description of astronomical tests; using it, one may cast this quantity as
\beq \Delta\left({M_G \over M_I}\right)_{\tt SEP} =  \eta\left({\Omega \over Mc^2}\right)~~, \label{eq:MgMi} \eeq
\noindent where $Mc^2$ is the total mass-energy of the test body, $\Omega $ its negative gravitational self-energy and $\eta$ a dimensionless constant for SEP violation \cite{Nordtvedt_1968a,Nordtvedt_1968b,Ken_LLR68}: it is expressed by a combination of the PPN parameters, so that in fully-conservative, Lorentz-invariant theories of gravity \cite{Will1,Will2005}, it reads $ \eta = 4\beta - \gamma -3$ (so that the values $ \beta = \gamma = 1$ characterizing GR yield $\eta = 0$).

The self energy of a body $B$ is given by
\beq \left({\Omega  \over Mc^2}\right)_B = - {G \over 2 M_B c^2}\int_B d^3{\bf x} d^3{\bf y} {\rho({\bf x})\rho({\bf y}) \over | {\bf x} - {\bf y}|}~~. \label{eq:omega} \eeq
\noindent A sphere with a radius $R$ and uniform density has $\Omega /Mc^2 = -3GM/5Rc^2 = -(3/10) (v_E/c)^2$, where $v_E$ is its escape velocity. A more realistic value may be obtained by numerically integrating the expression above using its known structural features: in the case of the Sun, this yields $(\Omega /Mc^2)_\odot \sim -3.52 \times 10^{-6}$ \cite{Ulrich_1982}, which should be compared with the typical magnitude $\sim 10^{-25}$ for laboratory sized bodies. Thus, while an experimental accuracy of $10^{-13}$ \cite{Adelberger_2001} is sufficient to significantly constraint violations of the WEP, it does not allow for a stringent test of the SEP --- hence the need for planetary-sized extended bodies, where the ratio Eq.~(\ref{eq:omega}) is much larger.

Several Solar System experiments have been suggested in order to probe the validity of the SEP \cite{Nordtvedt_1968a,Ken_LLR68,Nordtvedt_1970}, from lunar measurements to the study of the motion of Trojan asteroids (performed more than two decades ago \cite{Orellana_Vucetich_1988,Orellana_Vucetich_1993}) or the analysis of binary pulsar data \cite{Damour_Schafer_1991} --- which takes advantage of a strong (self-)gravity regime \cite{Damour_EFarese96a,Damour_EFarese96b}, albeit no sufficiently accurate measurements are yet available \cite{Wex_2001,Lorimer_Freire_2004}. Interplanetary spacecraft provide yet another testbed for the SEP \cite{Anderson_etal_1996,Anderson_Williams_2001}.

So far, the most competitive assessment of the validity of the SEP stems from the Earth-Moon-Sun system, through the analysis of lunar laser ranging (LLR) data \cite{Williams_Turyshev_Boggs_2004} yielding $\Delta (M_G/M_I)_{\tt SEP}=(-2.0\pm2.0)\times10^{-13}$ (from a general breaking of the EP of $\Delta (M_G/M_I)_{EP} =(-1.0\pm 1.4)\times 10^{-13}$) --- implying a SEP violation parameter $\eta=4\beta-\gamma-3= (4.4\pm4.5)\times 10^{-4}$.

\subsection{Local Lorentz Invariance (LLI)}
\label{LLI}

Invariance under Lorentz transformations states that the laws of physics are independent of the velocity of the frame. This is the basic tenet of S, as discussed before, and holds only locally in GR. Although current theories obey this symmetry, some results arising from string field theory hint that it may be spontaneously broken \cite{Kostelecky1989a,Kostelecky1989b}, due to open string interactions and its implications at low-energy physics. If so, many implications are expected: for instance, if the contribution of Lorentz-violating interactions to the vacuum energy is approximately half of the critical density, one expects that very weak tensor-mediated interactions arise in the range $ \sim 10^{-4}$~m \cite{Bertolami3}. Furthermore, these string interactions are the privileged contributors to the Lorentz violating terms of the mSME.

The effect of our velocity relative to a putative preferred reference frame may be phenomenologically described by considering a cosmological vector field that acquires a non-vanishing minima due to a spontaneous symmetry breaking induced by a suitable potential \cite{Phillips}; a model allowing for such a spontaneous breaking of LLI has been proposed \cite{Kostelecky4,Kostelecky5,Bluhm}, leading to interesting scenarios where the inverse square law for gravity is modified by the spacetime direction chosen by the vector field \cite{flight}.

Considerations on the dynamics of the renormalization group $\beta$-function of non-abelian gauge theories also hint that Lorentz invariance might be just a low-energy symmetry \cite{Nielsen}. Lorentz violation may also induce the breaking of conformal symmetry; together with inflation, this could explain the primordial magnetic fields needed to account for the observed galactic magnetic field \cite{Bertolami4}. A modified gravity-induced wave dispersion derived from a violation of Lorentz invariance could be probed by astrophysical observations of distant sources of gamma radiation \cite{Amelino1, Biller}.

A violation of this fundamental symmetry of GR is also possible with non-commutative field theories \cite{Carroll}, although it may hold (at least) at first non-trivial order in perturbation theory of the non-commutative parameter \cite{Bertolami15,Bertolami16,Imai,Conroy,Robbins}. Other theories that may entail a breaking of  Lorentz invariance include loop quantum gravity \cite{Gambini,Alfaro}, spacetime foam scenarios \cite{Garay,Ellis} and models exhibiting a spacetime variation of fundamental coupling constants \cite{Lehnerta,Lehnertb} (see Ref. \cite{Mattinglya} for a review of high-energy Lorentz symmetry breaking).

A violation of Lorentz invariance could break the fundamental CPT symmetry of local quantum field theories \cite{Kostelecky1995,Kostelecky1996} --- a prospect that can be tested in neutral-meson \cite{Colladaya,Colladayb} experiments, Penning-trap measurements \cite{Bluhm1997,Bluhm1998} and hydrogen-antihydrogen spectroscopy \cite{Bluhm2}. This CPT breaking could also be induced by non-linearities in quantum mechanics, perhaps stemming from a quantum theory of gravity; the latter possibility has been probed by the CPLEAR Collaboration \cite{Adler}. Whatever the cause, the spontaneous breaking of CPT symmetry  provides, along with the violation of the baryon number, an interesting mechanism for the generation of the observed baryon asymmetry in the Universe: after the  CPT and baryon number symmetries are broken in the early Universe, tensor-fermion interactions arising from string field theory give rise to a chemical potential that creates a baryon-antibaryon asymmetry in equilibrium \cite{Bertolami5}.

Modifications of the Michelson-Morley experiment using laser interferometry are very useful for testing Lorentz symmetry breaking, by comparing the velocity of light and the maximum attainable velocity of massive particles, $c_i$ --- with a current experimental constraint of $ |c^2/c_{i}^2 - 1| < 10^{-9}$ \cite{Brillet} (see Section \ref{exptestsSR}).

The more accurate Hughes-Drever experiment probe a possible time dependence of the quadrupole splitting of nuclear Zeeman levels along Earth's orbit \cite{Hughes,Drever}, yielding an impressive limit of $|c^2/c_{i}^2 - 1| < 3 \times 10^{-22}$ \cite{Lamoreaux} --- with a follow-up study showing that a gain of up to eight orders of magnitude in accuracy is possible \cite{Kostelecky3}.

As stated before, astronomical tests are best analyzed through the use of the PPN formalism, with the $\alpha_{3}$ parameter being related to violation of momentum conservation and the existence of a preferred reference frame ($\al_3 = 0$ in GR). The study of (millisecond) pulsars yields the extremely accurate limit $|\alpha_{3}| < 2.2 \times 10^{-20}$ \cite{Will2005, Bell,BellD}.

An analysis of the interaction between the most energetic cosmic-ray particles and the photons from the cosmic microwave background radiation has shown that the propagation of ultra-high-energy nucleons is limited by inelastic collisions with the latter, preventing particles with energies above $5 \times 10^{19}~\ev$ from reaching Earth from beyond $ 50--100 ~\mpc$ --- the so-called Greisen-Zatsepin-Kuzmin (GZK) cut-off \cite{Greisen1966,Zatsepin1966}. Events where the cosmic primaries have an estimated energy above the GZK cut-off where observed by different collaborations \cite{Hayashida,Takeda,Birda,Birdb,Birdc,Brooke,Efimov}; although the HI-RES (High Resolution Fly's Eye) \cite{Abbasi} and Auger \cite{Auger} collaborations results have been interpreted as being consistent with the validity of this cutoff, and hence of Lorentz symmetry.

Processes such as the resonant scattering reaction $p + \ga_{2.73K} \to \De_{1232}$ have been shown to be suppressed by energy-dependent effects arising from a small violations of Lorentz invariance \cite{Sato,Coleman1997,Coleman1999,Mestres}. This can be used to analyze the putative existence of events above the GZK cutoff, yielding the strongest constraint of $|c^2/c_{i}^2 - 1|  < 1.7 \times 10^{-25}$ \cite{Bertolami2,Bertolami6,Bertolami:2003yi}.

\subsection{Local Position Invariance (LPI)}
\label{LPI}

A violation of the LPI indicates that the rates of a free falling clock and one the surface of the Earth should differ. As the WEP and LLI principles of GR benefit of the stringent bounds addressed before, experiments on the universality of the gravitational red-shift primarily probe the validity of the LPI. This may be quantified by the parameter $\mu$ measuring the deviation in the relative shift in frequency $ \Delta \nu / \nu = (1 + \mu) U / c^2$ when compared with GR (where $\mu = 0$).

The already discussed Pound-Rebka experiment ({\it cf.} Eq. (\ref{eq:1.15})) yields $\mu \simeq 10^{-2}$; an accurate verification of the LPI was achieved through the comparison between hydrogen-maser frequencies on Earth and on a rocket flying to altitude of ten thousand kilometers \cite{Vessot}, leading to $ \vert \mu \vert < 2 \times 10^{-4}$. Further considerations allow for an improvement by two orders of magnitude, $ \mu < (0.1 \pm 1.4) \times 10^{-6}$ \cite{Ashby:2007zz}.

\subsection{The Pioneer and flyby anomalies}

Although not quite a direct test of SR or GR {\it per se}, the Pioneer and the flyby anomalies have arisen in the literature as phenomena that, at least at first look, did challenge the common wisdom about gravity. These unaccounted behaviour of spacecraft, derived from the analyses of tracking data, have led many theoreticians into the drawing board, with suggestions that these anomalies embodied new physical phenomena that could encompass a putative breaking of the basic tenets of Relativity.

The Pioneer anomaly stood out as an open question in physics for more than a decade: its existence was first discussed in 1998 \cite{Anderson1998}, when a JPL team showed that the deep tracking of the Pioneer 10 and 11 probes disagreed with the predictions of a detailed orbital determination model including GR and all relevant effects and ephemerides --- but was statistically consistent with a fit to the latter plus a constant sun-bound acceleration $a_P = (8.74 \pm 1.33) \times 10^{-10} ~ \rm{m/s^2}$ \cite{Anderson2002}. 

This anomalous behaviour was independently confirmed through alternative data analyses \cite{Markwardt2002,Toth2009,Levy2009}, with the first pair of studies allowing for a decreasing acceleration, instead of a constant one. Indeed, ten years ago it was pointed out that this was compatible with an exponentially decreasing acceleration with a time scale compatible with the decay rate of the plutonium present in the radiothermal generators (RTG) and powering the spacecraft. Nonetheless, and despite studies pointing at a conventional origin for the Pioneer anomaly \cite{Katz1999, Scheffer2003}, more specifically onboard thermal effects, this possibility was strongly rejected by the JPL team and explanations resorting to new physics appeared (see Refs. \cite{Bertolami2004, Reynaud2005, Moffat2006, Bertolami2007} and references therein). It was also shown that the most considered models for the mass distribution of the Kuiper Belt could not cause of the anomalous acceleration \cite{Bertolami2006} (see also Ref. \cite{Nieto2005}).

It was only in 2008 that a clear numerical indication that the Pioneer anomaly was of thermal origin did appear, with the radiation emitted from the RTGs and the main compartment providing the additional, decaying thrust that deviated the twin probes from its predicted trajectories \cite{thermal1}. This possibility gained strength with following independent studies \cite{thermal2,thermal3,thermal4,thermal5}, culminating in a recent study showing that the observed anomaly falls squarely into the predictions yielded by a model that also considers the reflection of the radiation on the parabolic dish of the high-gain antenna \cite{thermal6} --- a result confirmed by a subsequent study by other teams \cite{thermal7,thermal8}.

Thus, the Pioneer anomaly is no more, and now serves as a cautionary tale against the dangers of extrapolating poorly understood conventional effects as revolutionary evidence of deviations from SR and GR. With this is mind, the more recent flyby anomaly is viewed with added scepticism, although it has so far defied any conventional explanation.

\subsubsection{The flyby anomaly}

The flyby anomaly is an unexpected velocity change disclosed by the analysis of several Earth gravitational assist maneuvers of the Galileo, NEAR, Cassini and Rosetta spacecraft \cite{Anderson2008,Antreasian1998,AdlerUpdate}. Following flybys of the Galileo and Rosetta missions raised some some expectation of obtaining a confirmation of this phenomenon. However, these events yielded no further evidence of such a flyby anomaly (see Table~\ref{flyby_table}) --- in the case of the second Galileo flyby, due to the high uncertainty of the atmospheric drag, enhanced due to the very low perigee altitude of $\sim 300~km$.
 
With the exception of the Cassini spacecraft, the involved spacecraft had no Deep Space Network tracking during perigee passage, leading to an approximate four hour gap. The 10 s sampling interval for the remaining period produced a very coarse grained distribution of data points, disabling an accurate characterization of the effect in terms of an additional force affecting the spacecraft. Thus, the flyby anomaly is signaled by the inability to fit a single hyperbolic arc to the whole flyby maneuver: two distinct ``incoming'' and ``outgoing'' arcs have to be considered, with the small difference between them being interpreted as an additional boost $\Delta v$ at perigee.

Despite the difficult to assign a well defined value, an averaged acceleration of the order of $a_F \sim 10^{-4}~{\rm m/s^2}$ may be used as a figure of merit for the flyby anomaly \cite{Antreasian1998}. This figure allows for a comparison with several possible causes: Earth oblateness, other Solar System bodies, relativistic corrections, atmospheric drag, Earth albedo and infrared emissions, ocean or solid tides, solar pressure,  spacecraft charging, magnetic moments, solar wind, spin-rotation coupling \cite{Antreasian1998,Lammerzahl2006}, dark matter \cite{AdlerUpdate}, {\it etc}. (see Table \ref{error_sources_table}). 

Clearly, all these effects are much smaller than the considered value for $a_F$, with the exception of Earth oblateness. However, the accurate knowledge of the gravitational model of the Earth means that the origin of the flyby anomaly cannot be due to some minor deviation in the latter \cite{Antreasian1998}.

The empirical formula proposed in Ref. \cite{Anderson2008} is perhaps the most prominent attempt to account for the reported flyby anomalies; it proposes that the variation in magnitude and direction of the anomalous velocity change reflects the declinations of the incoming and outgoing asymptotic velocity vectors, $\delta_i$ and $\delta_o $, respectively:

\begin{equation}
	{\Delta V_\infty \over V_\infty} = {2\om_E R_E\over c} (\cos \delta_i - \cos \delta_o), \label{modelPRL}
\end{equation}
\noindent where $\om_E$ is the Earth's rotation velocity and $R_E$ its radius. This identification is suggestive, given its similarity with the term present in the outer metric due to a rotating body \cite{Ashby},

\begin{equation}
	ds^2 = \left(1 + 2{V - \Phi_0 \over c^2} \right)(c~dt)^2 - \left(1 - 2{V \over c^2} \right)(dr^2+r^2 d\Omega^2),
\end{equation}
with
\begin{equation}
	{\Phi_0 \over c^2} = {V_0 \over c^2} - {1 \over 2} \left( \omega_e R_e \over c \right)^2,
\end{equation}
where $d\Omega^2 = d\theta^2 + \sin^2\theta d\phi^2$, $V_0 $ is the Newtonian potential $V(r)$ at the equator.

However, any reasoning attempting to derive \eq{modelPRL} from GR is faulty, as all relativistic effects (embodied in the above metric) have been calculated to be much lower than the typical order of magnitude $a_F$ of the flyby anomaly --- namely those induced by the rotation of the Earth: the de Sitter precession effect and frame dragging.

Furthermore, the application of \eq{modelPRL} to the subsequent two flybys by the Rosetta probe in 2007 and 2008 predicted an anomalous increase in $V_\infty$ of respectively $0.98$ and $1.09$ mm/s \cite{Busack}, but the analysis of the tracking data was consistent with no flyby anomaly what so ever.

Similarly to what occurred with the Pioneer anomaly, a conventional explanation for the flyby anomaly should not be dismissed: indeed, some yet unmodelled aspect of the affected spacecrafts could lead to the observed anomalous $\Delta v$; if this is the case, the widely different designs and gravitational assists of the studied spacecrafts would naturally lead to the variations of the latter.

The opposite possibility might be more enticing, namely that the flyby anomaly is the signature of new or ``exotic" physics at play. Its confirmation as a new  physical force would have implications to a wide range of phenomena such as planetary orbits, and potentially lead to deepen our understanding of gravity. However, no clear cut fundamental motivation exists for such a short ranged force (see Refs. \cite{Lammerzahl2006} and \cite{HEO} for an overview of some proposed physical mechanisms).

In order to settle the issue, a clear cut confirmation of this effect is mandatory. Given the sparse number of gravitational assists available, a recent proposal \cite{HEO} has suggested that a thorough characterization of the flyby anomaly could be achieved by studying the behaviour of a spacecraft in a highly elliptic orbit, such that the velocity and altitude at perigee is similar to the reported values depicted in Table~\ref{flyby_table}. Such a mission could come at a low cost and would provide the desired repetition of flybys; a detailed study of its design features ({\it e.g.} thermal modelling, atmospheric drag) would allow for a clear discrimination of competing perturbations, and the use of Global Navigation Satellite System (GNSS) tracking would provide the required tracking accuracy.

This concept could be realized via a dedicated small or micro-satellite, or as an add-on to an existing mission \cite{HEO}. The STE-QUEST mission, currently under consideration by the European Space Agency, could provide the latter, given its highly elliptic orbit and use of GNSS precise orbit determination \cite{QUEST_Flyby}. 

\begin{table}
\begin{center}
\caption{\label{flyby_table}Summary of orbital parameters of the considered Earth flybys.}
\begin{tabular}{@{}ccccccc}
					\hline\hline

Mission	& Date	& $e$	& Perigee		& $v_\infty$		& $\Delta v_\infty$	& $\Delta v_\infty / v_\infty$ \\
					&		&		& $({\rm km})$	& $({\rm km/s})$ 	& $({\rm mm/s})$ 	& $(10^{-6})$ \\
					\hline
			Galileo	& 1990	& $2.47$& $959.9$		& $8.949$			& $3.92 \pm 0.08$	& $0.438$  \\  
			Galileo	& 1992	& $3.32$& $303.1$		& $8.877$			& $\sim 0 $		& $-0.518$ \\  
			NEAR	& 1998	& $1.81$& $538.8$		& $6.851$			& $13.46 \pm 0.13$	& $1.96$   \\
			Cassini	& 1999	& $5.8$	& $1173$		& $16.01$			& $-2 \pm 1$		& $-0.125$ \\  
			Rosetta	& 2005	& $1.327$& $1954$		& $3.863$			& $1.80 \pm 0.05$	& $0.466$  \\
			MESSENGER& 2005	& 1.360		& $2347$		& $4.056$			& $0.02 \pm 0.01$	& $0.0049$ \\
			Rosetta	& 2007	& 3.562		& $ 5322 $	& 9.36					& $\sim 0$			& - \\
			Rosetta	& 2009	& 2.956		& $2483$		& 9.38				& $\sim 0$			& - \\

\hline\hline

\end{tabular}
\end{center}
\end{table}


\begin{table}
\begin{center}
\caption{\label{error_sources_table}List of orders of magnitude of possible error sources during Earth flybys.}
\begin{tabular}{@{}cc}

\hline\hline
Effect				& Order of Magnitude \\
								& $({\rm m/s^2})$ \\
		\hline											
			Earth oblateness		& $10^{-2}$	\\
			Other Solar System bodies & $10^{-5}$	\\
			Relativistic effects		& $10^{-7}$	\\
			Atmospheric drag	& $10^{-7}$	\\
			Ocean and Earth tides	& $10^{-7}$	\\
			Solar pressure		& $10^{-7}$	\\
			Earth infrared		& $10^{-7}$	\\
			Spacecraft charge	& $10^{-8}$	\\
			Earth albedo		& $10^{-9}$	\\
			Solar wind			& $10^{-9}$	\\
			Magnetic moment		& $10^{-15}$	\\

\hline\hline

\end{tabular}
\end{center}
\end{table}

\subsection{Conclusion}
\label{tests}

As seen in the preceding sections, all of the available constraints on the validity of the founding principles of SR and GR have so far failed to crack any faults in these century-old theories, which thus remains the standard against all competitors so far.

The available experimental data fit quite well with GR, while allowing for the existence of putative extensions, provided any new effects are small at the post-Newtonian scale \cite{Will1}. However, despite its impressive experimental success, GR cannot be regarded as a fully satisfactory theory, given its inadequacy in what concerns issues such as the existence of singularities, the Cosmological Constant problem (see Ref. \cite{OBCC} and refs. therein) and the incompatibility with existing quantization schemes.

At the largest scales, GR is compatible with cosmological data if and only if dark matter dominates at galactic and clusters scales, while the dynamics of the accelerating expansion of the Universe is controlled by dark energy.

At a more conceptual level, it has been recently suggested that gravity, and GR in particular, is an emerging property arising from more fundamental tenets such as the holographic principle and Bekenstein's entropy-energy limit \cite{Verlinde}.

This perspective leads to new challenges and may imply, for instance, that the WEP might be violated whether space-times admits a phase-space non-commutative geometry \cite{Bastos}.

\bibliographystyle{unsrt}
\bibliography{handbook}

\end{document}